\documentclass[a4paper,11pt]{article}
\usepackage{jcappub}
\usepackage{lineno}
\title{\boldmath Low-energy threshold demonstration for dark matter searches in TREX-DM with an \texorpdfstring{$^{37}$Ar}{37Ar} source produced at CNA HiSPANoS}

\author[a]{J. Castel,}
\author[a]{S. Cebrián,}
\author[a]{T. Dafni,}
\author[a]{D. Díez-Ibáñez,}
\author[a]{A. Ezquerro,}
\author[b,c]{B. Fernández,}
\author[a]{J. Galán,}
\author[a]{J.A. García,}
\author[b,c]{C. Guerrero,}
\author[a]{I.G. Irastorza,}
\author[a]{G. Luzón,}
\author[a]{C. Margalejo,}
\author[a]{H. Mirallas,}
\author[a]{L. Obis,}
\author[a]{A. Ortiz de Solórzano,}
\author[a,1]{O. Pérez\note{Corresponding author.},}
\author[a]{J. Porrón,}
\author[a]{M. J. Puyuelo}
\author[a]{and A. Quintana}

\affiliation[a]{Centro de Astropartículas y Física de Altas Energías, Universidad de Zaragoza (CAPA),\\
50009 Zaragoza, Spain}
\affiliation[b]{Centro Nacional de Aceleradores (CNA) (US-Junta de Andalucía-CSIC),\\
41092 Sevilla, Spain}
\affiliation[c]{Dpto. Física Atómica, Molecular y Nuclear, Universidad de Sevilla, \\41012 Sevilla, Spain }

\emailAdd{oscarperlaz@unizar.es}

\abstract{We report on the successful implementation of an $^{37}$Ar calibration source in the TREX-DM detector, a high-pressure time projection chamber designed for low-mass dark matter searches. The $^{37}$Ar source was produced through fast neutron activation of CaO powder at the HiSPANoS facility of Centro Nacional de Aceleradores (CNA) in Spain, yielding $O(1)$~kBq of activity. Using a novel combined GEM-Micromegas readout system, we successfully detected both characteristic emissions from $^{37}$Ar decay (2.82~keV and 270~eV) and achieved unprecedented energy threshold performance in TREX-DM, approaching the single-electron ionization energy of argon.}

\begin{document}
\maketitle
\flushbottom

\section{Introduction}
\label{sec:intro}

Dark matter is a form of non-luminous, non-baryonic matter that makes up approximately 26\% of the mass-energy budget of the Universe~\cite{planck_2020}. Among all the particles proposed to constitute dark matter, Weakly Interacting Massive Particles (WIMPs) stand out as one of the leading candidates. However, the traditional mass range $m\sim 10\mathrm{~GeV/c}^2 - 1\mathrm{~TeV/c}^2$ motivated by the so-called "WIMP miracle"~\cite{WIMPs_lecture_notes_Feng_2022} has been increasingly constrained by numerous direct search experiments~\cite{Billard_2022} and by collider searches at facilities such as the LHC~\cite{DM_colliders_2018}. As a result, WIMP searches are gradually broadening their focus to include the low-mass region~\cite{Battaglieri_2017} ($< 10$~GeV/c$^2$), which remains the least explored region of the parameter space. This shift has been possible by significant improvements in detector sensitivity, which have allowed experiments to achieve very low energy thresholds (sub keV$_{ee}$)~\cite{Essig_2022, Diamond_2025}.

TREX-DM~\cite{TREX-DM_2016, TREX-DM_Bckg_Assessment_2018, TREX-DM_2024} is a high-pressure time projection chamber (TPC) designed to search for low-mass WIMPs using argon- or neon-based gas mixtures. The selection of light noble gases is motivated by their enhanced sensitivity to low-energy nuclear recoils characteristic of WIMP interactions, while operation at high pressure increases the target mass density and detection efficiency.

The detector consists of a cylindrical vessel made of radiopure copper, with a diameter of 0.5~m, a length of 0.5~m, and a wall thickness of 6~cm. These dimensions are determined by the requirements to withstand pressures up to 10~bar(a) while serving as the innermost component of the shielding system. The vessel houses a total active volume of 20~L, which corresponds to approximately 0.3~kg of argon or 0.16~kg of neon at 10~bar.

A central cathode, connected to high voltage through a custom-made feedthrough, divides the vessel into two symmetric sensitive volumes. Each volume includes a 16-cm-long drift region defined by a set of copper strips printed on a Kapton substrate and supported by four PTFE walls. This configuration ensures a uniform electric field throughout the active volume.

The readout system is made up of two $25\times 25$~cm$^2$ Microbulk Micromegas detectors~\cite{Giomataris_1995, Andriamonje_2010}, placed at opposite ends of the active volumes. Each detector is segmented into 256 channels in the X direction and 256 in the Y direction, with a strip pitch of approximately 1~mm. This layout provides excellent spatial resolution~\cite{Derre_2001} and background discrimination capabilities. Micromegas technology is particularly well-suited for rare event searches due to its intrinsically low radioactivity~\cite{Cebrian_2010} and the potential to achieve low-energy thresholds~\cite{Derre_2000}. Furthermore, recent work has shown that stacking a GEM preamplification stage on top of a microbulk Micromegas, such as the ones used in TREX-DM, can provide extra preamplification factors in the range $20-100$~\cite{GEM_2025}, contingent on gas pressure.

The experiment is located at the Laboratorio Subterráneo de Canfranc (LSC) under the Spanish Pyrenees, where the rock overburden (2450 m.w.e.) provides natural shielding against cosmic muons and associated secondary particles.

The energy threshold and resolution directly impact the sensitivity reach for low-mass WIMP searches, where nuclear recoil energies can be as low as a few keV or below~\cite{Lewin_1996}. Understanding detector response and achieving precise event reconstruction at such low energies is critical for distinguishing potential WIMP signals from background events. Most commonly used external calibration sources suffer from detector self-shielding and limited penetration depth, resulting in non-uniform event distributions throughout the active volume. Furthermore, most standard calibration sources lack discrete emissions in the sub-keV energy range. For this purpose, $^{37}$Ar has emerged as an ideal calibration source. This gaseous radioisotope ($t_{1/2}=35.04$~d) decays to $^{37}$Cl via electron capture ($Q=813.9$~keV), creating vacancies in the atomic electron shells. The subsequent atomic relaxation process produces discrete, monoenergetic emissions: 2.82~keV from K-shell capture (90\% probability) and 0.27~keV from L-shell capture (9\% probability)~\cite{lnhb_table_radionucleides}. The gaseous nature of $^{37}$Ar ensures a homogeneous distribution across the sensing planes, eliminating spatial biases inherent to external point sources.

This work presents the development and implementation of a $^{37}$Ar calibration source for the TREX-DM detector, demonstrating its capability to achieve exceptionally low energy thresholds. In Section~\ref{sec:production_irradiation}, we outline the chosen production method and the irradiation process carried out at Centro Nacional de Aceleradores (CNA) in Sevilla. Section~\ref{sec:calibration} analyzes the low-energy calibration spectrum obtained with the TREX-DM detector, where an exceptional energy threshold of $O(10)$~eV was achieved thanks to the gain provided by the combined GEM-Micromegas structure. Finally, Section~\ref{sec:conclusions} summarizes our findings and discusses their implications for low-mass WIMP searches.

\section{Production Method and Irradiation at CNA HiSPANoS}
\label{sec:production_irradiation}

\subsection{Production Method and Setup}
\label{sec:production}

Several production routes have been investigated for the generation of $^{37}$Ar: 

\begin{itemize}
    \item Proton bombardment through the reaction $^{37}$Cl(p,n)$^{37}$Ar, as explored in~\cite{Kishore_1975}. The cross section of this reaction was studied in~\cite{Weber_1985}.
    \item Activation of $^{36}$Ar with thermal neutrons via the reaction $^{36}$Ar(n,$\gamma$)$^{37}$Ar. This can be achieved using atmospheric argon~\cite{Sangiorgio_2013} or, as implemented by the XENON collaboration, using argon enriched in $^{36}$Ar~\cite{XENON1T_Ar37_2022}. The enrichment minimizes the production of unwanted isotopes such as $^{39}$Ar ($t_{1/2}=269$~years) and $^{41}$Ar ($t_{1/2}=110$~minutes), while maximizing the yield of $^{37}$Ar. The cross section of the reaction for thermal neutrons is in the range $1-10$~b~\cite{nndc_ENDF_2011}.
    \item Neutron activation of calcium oxide (CaO) via $^{40}$Ca(n,$\alpha$)$^{37}$Ar. In this case, the use of thermal neutrons is possible (cross section in the range $0.1-10$~mb~\cite{nndc_ENDF_2011}), but the cross section for fast neutrons (peaking at approximately $200$~mb for 5-7~MeV neutrons~\cite{nndc_ENDF_2011}) dominates the reaction~\cite{Kelly_2018}. This represents the most common method for the production of $^{37}$Ar and it has demonstrated success across multiple experimental contexts, including dual-phase xenon TPCs~\cite{Boulton_2017} and gas-based TPCs such as the one used by NEWS-G~\cite{NEWS_Ar37_2014}.
\end{itemize}

For implementation in TREX-DM, the CaO irradiation method was selected based on the ready availability of CaO powder and the possibility to access a suitable fast neutron facility, thanks to our collaboration with CNA. 

\begin{figure}[htbp]
\centering
\includegraphics[width=1.0\textwidth]{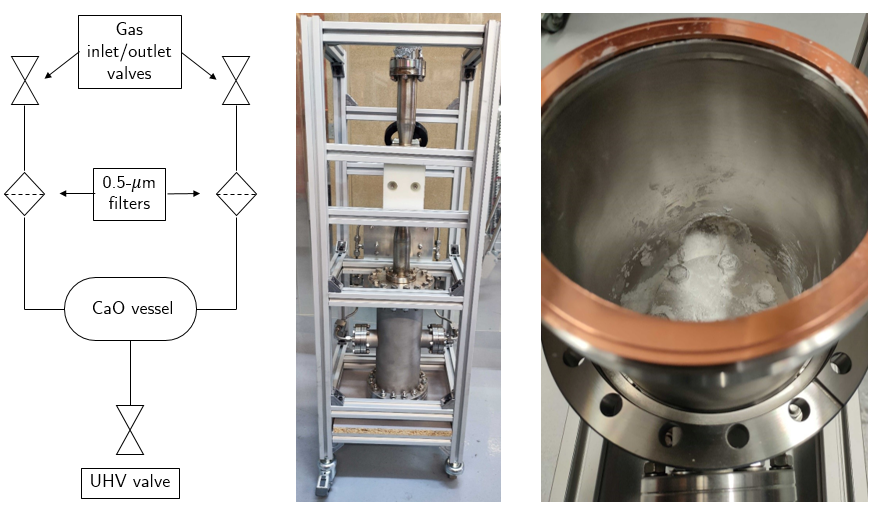}
\caption{Left: P\&ID diagram with the different components of the system: stainless-steel cross used as a container for the CaO powder, particulate filters to keep the powder isolated, valves for source injection, and UHV valve through which the system is pumped before irradiation. Center: image of the actual set-up where the external support structure is visible. Right: vessel filled with 0.5~kg of powder. It can be observed that the upper part and the side filters are left unobstructed so that the $^{37}$Ar liberated from the CaO can easily flow into the gas system. \label{fig:TREX-DM_setup}}
\end{figure}

The set-up for irradiation comprised a stainless-steel cross-shaped vessel (CF DN40 - CF DN100) filled with 0.5~kg of high-purity CaO powder (99.9\% purity from Sigma Aldrich). Two 0.5-$\mu$m particulate filters (Swagelok) were installed on the vessel sides to prevent powder contamination during gas transfer operations. An Ultra High Vacuum (UHV) valve mounted on the top part of the cross ensures appropriate vacuum levels can be achieved and maintained, and a 4-$\mu$m filter is placed in the UHV line to protect the turbopump during evacuation. The complete system underwent comprehensive leak testing up to 10~bar. A custom support structure was fabricated to ensure mechanical stability during transport between facilities and handling operations. A P\&ID sketch and some images of the set-up can be seen in Figure~\ref{fig:TREX-DM_setup}.

\subsection{Irradiation Process at CNA HiSPANoS}
\label{sec:irradiation}

\begin{figure}[htbp]
\centering
\includegraphics[width=0.64\textwidth]{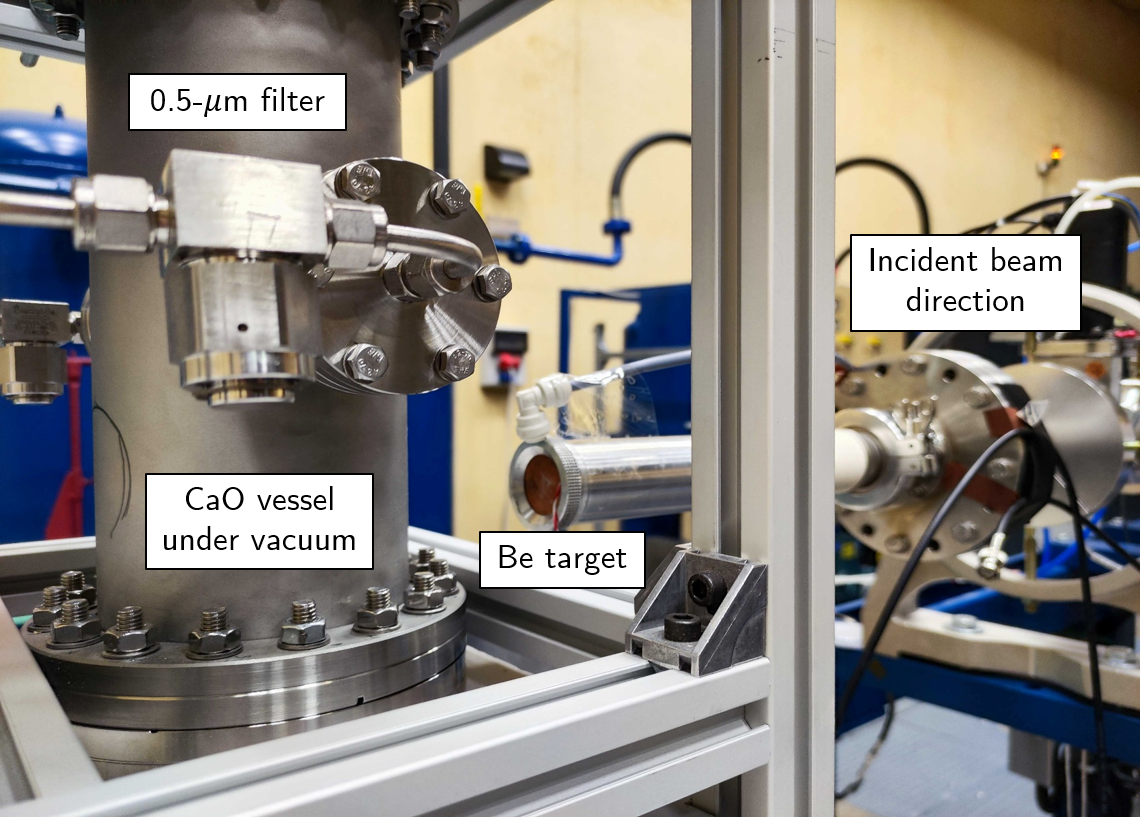}
\caption{Irradiation setup at CNA irradiation hall. The distance beam-vessel is 75~mm. \label{fig:irradiation_setup}}
\end{figure}

The neutron irradiation was conducted at the Centro Nacional de Aceleradores (CNA) in Sevilla using their HiSPANoS neutron beam facility~\cite{Gomez-Camacho_2021,HiSPANoS_2024}. At HiSPANoS, continuous or pulsed beams of thermal, epithermal or fast neutrons can be produced through the interaction of proton or deuteron beams of up to 6~MeV with D, Li or Be targets. For the purpose of efficient production of $^{37}$Ar, the neutron yield and energy needed to be maximized. Hence, $\sim3\times10^{10}$~n/sr/s fast neutrons were produced bombarding an air-cooled beryllium target with a 5~MeV deuteron beam at 1.2~$\mu$A. The irradiation consisted of approximately 7~h of beam exposure, with the CaO vessel kept under vacuum ($\sim 10^{-3}$~mbar) and positioned at 75~mm from the beryllium target (see Figure~\ref{fig:irradiation_setup}).

In addition to $^{37}$Ar, natural calcium irradiation simultaneously produces $^{42}$K via the reaction $^{42}$Ca(n,p)$^{42}$K, which serves as a valuable production monitor through its characteristic 1525~keV gamma emission ($t_{1/2} = 12.4$~h). Although the natural abundance of $^{42}$Ca (0.647\%) is significantly lower than $^{40}$Ca (96.941\%), the reaction still provides practical means to assess the success of the reaction when direct detection of $^{37}$Ar is not feasible. To track this gamma emission, two scintillation detectors (a 3~in $\times$ 3~in NaI and a 1.5~in $\times$ 1.5~in LaBr$_3$) were used after the irradiation procedure was completed, previously calibrated with standard sources ($^{137}$Cs at 662~keV, $^{60}$Co at 1173 and 1332~keV). Given the presence of unidentified spectral components coming from the activation of materials such as stainless steel, a precise determination of $^{42}$K activity immediately after irradiation proved to be difficult. Therefore, only a rough estimate of $^{42}$K activity was obtained. This was done independently with each detector, producing compatible results (normalized by detector area). Using the resulting $^{42}$K activity, the $^{37}$Ar activity was extrapolated to be $O(1)$~kBq (see~\cite{thesis_perez_2025} for more details on how to estimate the $^{37}$Ar activity).

Following neutron irradiation, the radioactivity of the vessel was monitored using a dosimeter until ambient background levels were achieved ($\lesssim0.2$~$\mu$Sv/h), which required 4-5 days. The source was then transported directly to LSC, with a total time of two weeks between irradiation and injection into TREX-DM. This delay resulted in approximately 25\% activity loss due to radioactive decay, a negligible amount given the initial kBq-level production.

\section{Calibration in TREX-DM}
\label{sec:calibration}

The $^{37}$Ar calibration measurement was conducted with the TREX-DM detector equipped with a fully operational GEM-Micromegas readout system such as the one presented in our previous work~\cite{GEM_2025}. The gas mixture used in this calibration is Ar-1\%iC$_{4}$H$_{10}$ and the detector pressure $\sim$ 1~bar(a). To process and analyze the data, we used REST-for-Physics~\cite{REST_2022}, a ROOT-based software framework developed for the analysis of data from rare event search experiments.

The injection of $^{37}$Ar into the TREX-DM chamber was performed through a controlled multi-cycle process. The irradiated container was first filled with Ar-1\%iC$_{4}$H$_{10}$ gas mixture at 2~bar, then connected to the detector gas system maintained at 0.9~bar. The connection to the gas system was made by passing the gas through a moisture filter and an oxygen filter, with the aim of ensuring that the injection of the source did not introduce impurities that could worsen the detector's performance.  After pressure equilibration, the chamber was isolated and the container refilled. This procedure was repeated four times, with each cycle extracting approximately 50\% of the remaining $^{37}$Ar activity and increasing the chamber pressure by $\sim$ 0.04~bar per cycle.

\subsection{Low-Energy Spectrum}
\label{sec:calibration_spectrum}

Once the source was injected into the detector volume, data acquisition was performed at high-voltage settings ($V_{\mathrm{mesh}}=295$~V, $V_{\mathrm{GEM}}= 280$~V). The choice of voltages corresponds to the highest stable operating point achieved with the TREX-DM detector, with the goal of maximizing gain and, therefore, low-energy sensitivity.

\begin{figure}[htbp]
\centering
\includegraphics[width=1.0\textwidth]{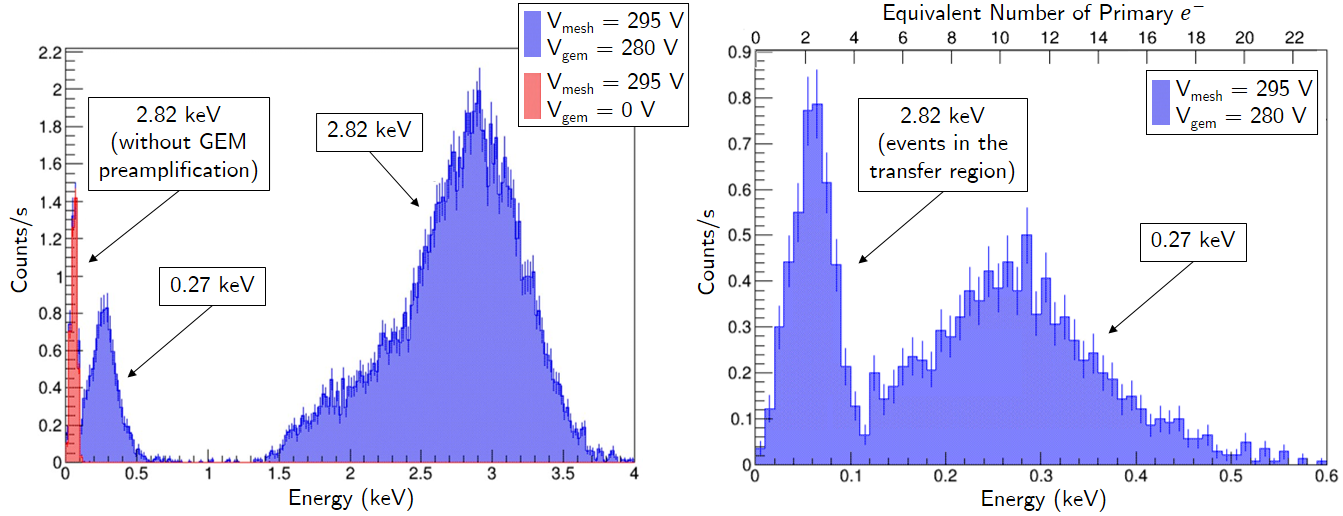}
\caption{Complete $^{37}$Ar energy spectrum obtained with TREX-DM. Total calibration time is $\sim$ 3 min due to the high rate of the source. Left panel shows the full spectrum with both characteristic peaks: 2.82~keV from K-shell electron capture and 270~eV from L-shell capture. The blue curve corresponds to the GEM-Micromegas configuration with both amplification stages active, while the red curve shows the spectrum with only the Micromegas detector powered on. Right panel shows the zoomed view of the low-energy region, showing a threshold of $O(10)$~eV in equivalent energy. The $x$ axis is also represented as the equivalent number of primary electrons, defined as the equivalent energy divided by the average energy needed to produce an electron-ion pair ($\approx 26$~eV in argon). \label{fig:Ar37_spectrum}}
\end{figure}

Figure~\ref{fig:Ar37_spectrum} presents the complete $^{37}$Ar energy spectrum obtained with TREX-DM. The blue curve corresponds to the full GEM-MM configuration with both amplification stages active, while the red curve shows the spectrum acquired with only the Micromegas stage operational ($V_{\mathrm{GEM}}= 0$~V). The difference between these configurations demonstrates the significant preamplification provided by the GEM foil. The total rate of the source was $\approx 3$~kBq, in line with the estimate obtained from $^{42}$K monitoring, and the total run time was $\sim 3$~min.

The spectrum clearly reveals both characteristic $^{37}$Ar emissions: the 2.82~keV peak from K-shell electron capture (used to calibrate the spectrum) and, remarkably, the 270~eV peak from L-shell capture. The ratio of events between the K-shell peak and the L-shell peak is $10:1$, in accordance with the expected decay probabilities. The observation of this ultra-low-energy peak represents a significant achievement for TREX-DM, demonstrating sensitivity in the energy region critical for light WIMP searches.

The low-energy portion of the spectrum (right panel of Figure~\ref{fig:Ar37_spectrum}) shows that events are detected down to equivalent energies of $O(10)$~eV (see~\ref{sec:calibration_energy_threshold}), approaching the single-electron ionization energy of argon. This threshold performance validates the suitability of the GEM-MM readout technology for rare event search applications.

Analysis of the additional preamplification factor provided by the GEM shows close agreement with test bench measurements: comparing the 2.82~keV peak position in the Micromegas-only configuration (equivalent energy $\sim$ 50~eV) with that of the GEM-MM configuration yields a preamplification factor of $50-60$, consistent with our previous characterization studies~\cite{GEM_2025}.

Lastly, the values for the drift field and transfer field used here were taken from our optimization studies carried out in~\cite{GEM_2025}. Several runs were taken to ensure that extraction efficiency of the GEM foil and transparency of the Micromegas structure were optimal with these settings. What remains to be studied is the possible field dependence of the extraction for few-electron events specifically; addressing this would require a dedicated study of the transfer field or simulations of the electron transport at the few-electron level, and is left for future work.

\subsection{Spatial Homogeneity Assessment}
\label{sec:calibration_homogeneity}

\begin{figure}[htbp]
\centering
\includegraphics[width=1.0\textwidth]{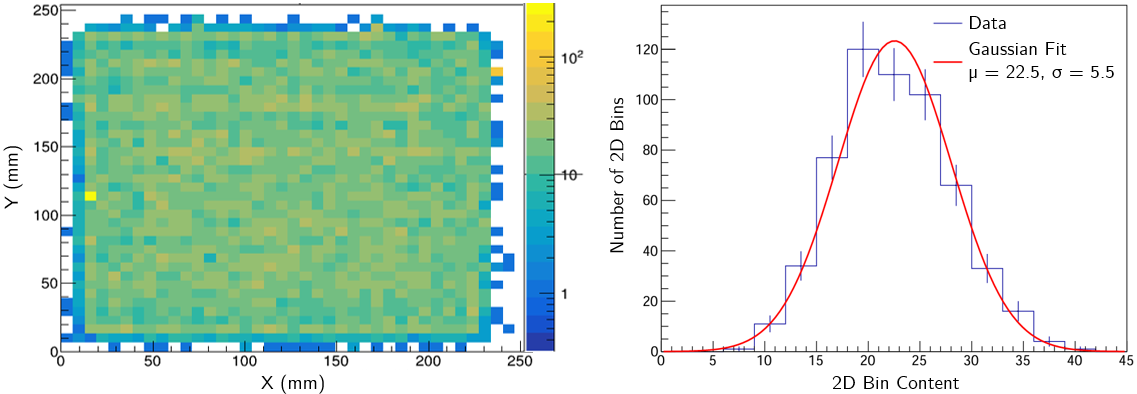}
\caption{Spatial distribution analysis of $^{37}$Ar events in TREX-DM. Left: 2D hitmap qualitatively illustrating spatial uniformity across the whole detector area. Right: 1D projection of bin counts from the inner 20$\times$20~cm$^2$ fiducial region, together with a Gaussian fit (red line). The standard deviation approximately follows the expected $\sigma=\sqrt{\mu}$. \label{fig:Ar37_hitmap}}
\end{figure}

The spatial distribution of $^{37}$Ar events throughout the detector active volume was evaluated through analysis of the reconstructed 2D positions. Figure~\ref{fig:Ar37_hitmap} shows the hitmap corresponding to the integrated calibration runs, demonstrating excellent spatial uniformity across the detector area.

Quantitative assessment of the spatial homogeneity was performed by analyzing the statistical distribution of event counts within a 20$\times$20~cm$^2$ fiducial region, excluding potential edge effects and pixels containing an excess of noisy events (such as the ones at $X\approx 15$~mm, $Y\approx 115$~mm and $X\approx 235$~mm, $Y\approx 205$~mm). The bin count distribution follows the expected Poisson statistics, with the 1D projection (right part of Figure~\ref{fig:Ar37_hitmap}) showing excellent agreement with a Gaussian fit of mean $\mu$ and standard deviation $\sigma$. The extracted standard deviation matches the expected relationship $\sigma=\sqrt{\mu}$, confirming the proper uniformity of the detector response. We emphasize that this one-dimensional test statistic is used only as a complementary check: it verifies that bin contents are statistically consistent with a Gaussian distribution, but it does not by itself prove spatial homogeneity, as this is primarily established visually from the 2D map.

The absence of visible dead regions or response variations across the detector area is crucial, as non-uniform detector response could introduce systematic biases in background discrimination and signal identification analyses.

\subsection{Energy Threshold Determination}
\label{sec:calibration_energy_threshold}

\begin{figure}[htbp]
\centering
\includegraphics[width=1.0\textwidth]{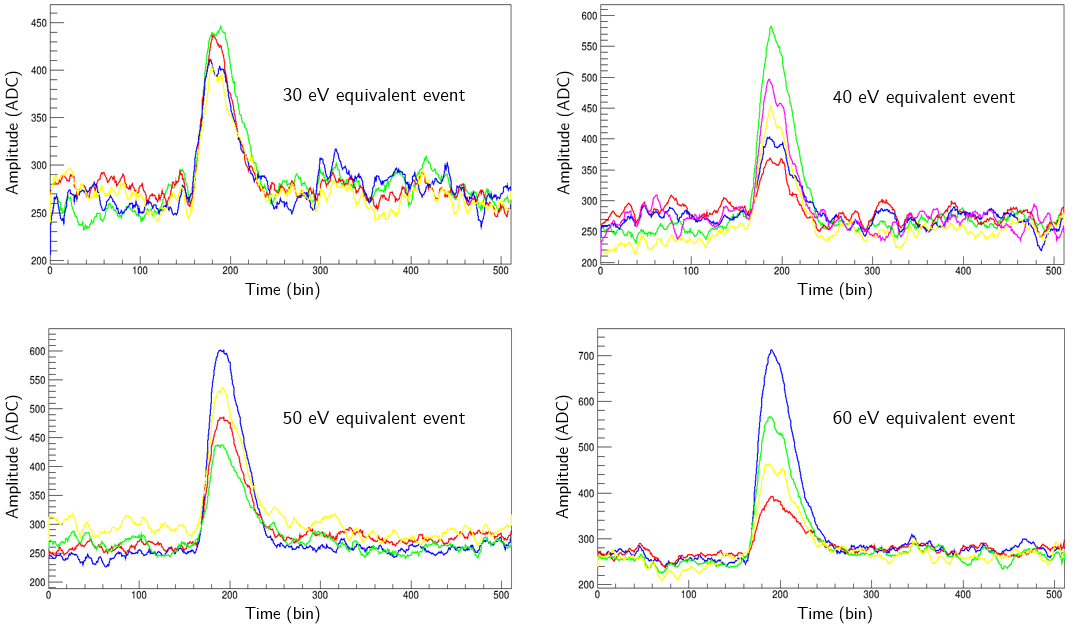}
\caption{Some representative low-energy events recorded by TREX-DM using the GEM-MM readout system during $^{37}$Ar calibration under high-gain conditions. The different colors correspond to signals from individual detector strips within the same event. The signals are clearly distinguishable from baseline noise, even for events with equivalent energies of 30~eV.
 \label{fig:Ar37_low_energy_events}}
\end{figure}

Looking at the spectrum on the right side of Figure~\ref{fig:Ar37_spectrum}, we observe a population of events with energies below 100~eV. We can also see that there is a complete overlap between these events and the peak from the run with the GEM turned off (in red on the left side of Figure~\ref{fig:Ar37_spectrum}). This leads us to conclude that these events originate from 2.82 keV $^{37}$Ar decays occurring in the transfer region, the gap between the GEM and Micromegas stages where ionization electrons receive amplification only from the Micromegas. Since these events experience reduced total gain compared to drift region events (which benefit from both GEM and Micromegas amplification), they appear at lower equivalent energies in the spectrum. This demonstrates that our detection capability would extend to such low-energy events if there was a drift region population at these energies, given that our electronics threshold operates well below 100~eV. We stress that these events do not correspond to true $\sim$ 20 eV depositions, but their pulse amplitude is equivalent to the one produced by a few-electron signal. To highlight this, we have also expressed the $x$ axis of the plot in terms of the equivalent number of primary electrons, defined as the equivalent energy divided over the average energy needed to produce an electron-ion pair ($\approx 26$~eV in argon). In what follows, we will work with energy units, but noting that it is an equivalent energy and that the results can also be expressed as the equivalent number of primary electrons. Analysis of representative low-energy events (illustrated in Figure~\ref{fig:Ar37_low_energy_events}) shows that these signals represent genuine physical events well above baseline noise levels, with reliable event identification reaching equivalent energies as low as 30~eV.

The determination of the detector energy threshold was performed through systematic analysis of the shape of this low-energy transfer region peak as a function of amplification voltage. The threshold manifests as a low-energy cut-off in the peak profile, which becomes less pronounced as the detector gain increases. This is modeled using a cropped-peak fit consisting of a Gaussian peak multiplied by a smoothed step function:

\begin{equation}
\label{eq:fit_energy_threshold}
f(E) = A \ \exp\left( -\frac{(E - \mu)^2}{2\sigma^2} \right) \times \frac{1}{2} \left[ 1 + \mathrm{erf} \left( \frac{E - E_{\text{thr}}}{\sqrt{2} \, \delta} \right) \right] \,,
\end{equation}

\noindent where $A$, $\mu$ and $\sigma$ are the amplitude, mean position and standard deviation of the Gaussian, $E_\mathrm{thr}$ is the energy threshold, and $\delta$ is a parameter that regulates the smoothness of the transition in the step function, which has been implemented through the error function. All five parameters are left free in the fit. In particular, the energy threshold $E_\mathrm{thr}$ is the key parameter that we seek to estimate.

\begin{figure}[htbp]
\centering
\includegraphics[width=1.0\textwidth]{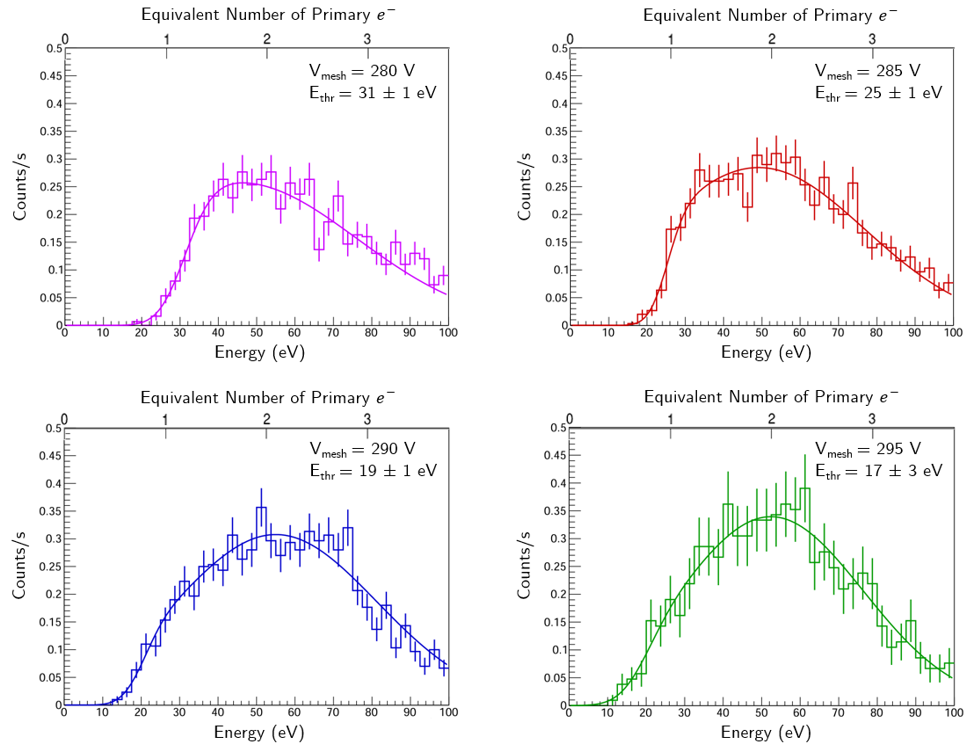}
\caption{Fit to the cropped-peak model from Equation~\ref{eq:fit_energy_threshold} (solid curve) for different mesh voltages (280, 285, 290, and 295~V) together with the experimental low-energy spectra, shown with error bars. Fitted energy threshold values and statistical uncertainties are shown for each voltage. The threshold improves progressively with increasing voltage, validating the expected behavior. \label{fig:Ar37_energy_threshold_fits}}
\end{figure}

\begin{figure}[htbp]
\centering
\includegraphics[width=1.0\textwidth]{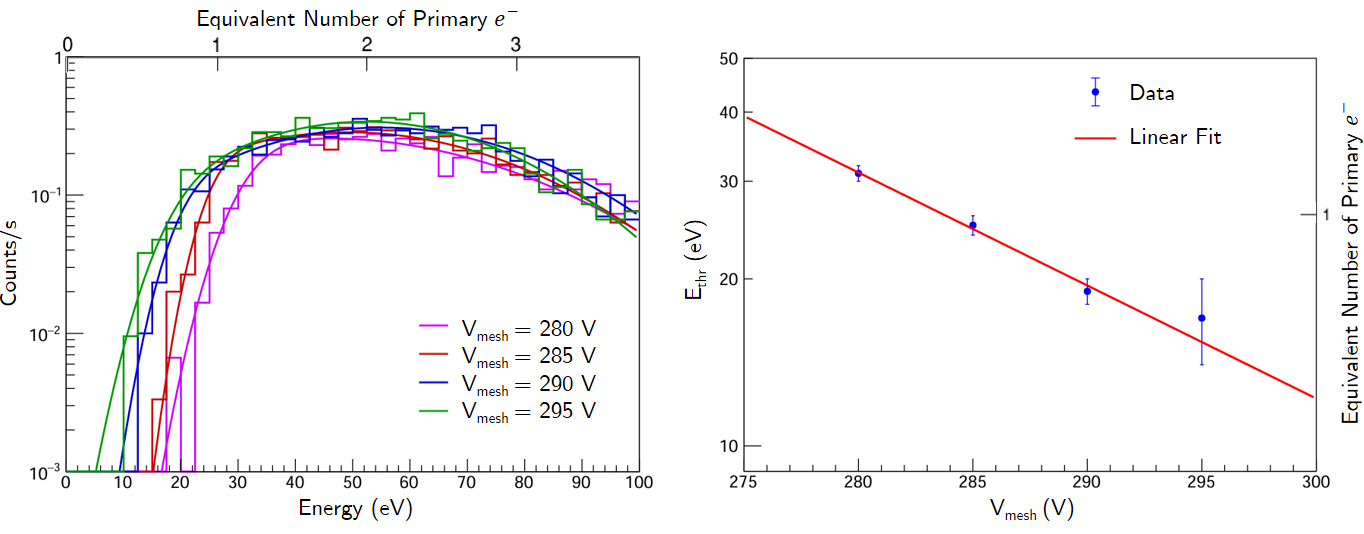}
\caption{Left: overlaid energy spectra for four different mesh voltage settings (280, 285, 290, and 295~V) with corresponding fits, displayed on logarithmic scale to emphasize threshold behavior. Right: energy threshold in logarithmic scale as a function of mesh voltage, together with a fit to the expected linear dependence $\ln(E_\mathrm{thr})\propto V_\mathrm{mesh}$. The slight deviation observed at the highest gain (295~V) remains within statistical uncertainty and reflects reduced fitting sensitivity when the Gaussian peak profile becomes nearly complete. The achieved thresholds confirm the exceptional low-energy sensitivity of the TREX-DM detector system. \label{fig:Ar37_energy_threshold_determination}}
\end{figure}

Figure~\ref{fig:Ar37_energy_threshold_fits} presents the threshold analysis results for four different mesh voltage settings (280, 285, 290, and 295~V), with GEM voltage fixed. Thresholds range from approximately 60~eV at 280~V to 20~eV at 295~V. The systematic decrease in energy threshold with increasing voltage is more clearly seen in Figure~\ref{fig:Ar37_energy_threshold_determination} (left side), confirming the expected detector behavior. The voltage dependence of the energy threshold follows the expected relationship $E_\mathrm{thr}\sim 1/G$, where $G$ represents the amplification gain of the system. Since $G\sim \exp(aV_\mathrm{mesh})$, with $a$ being some constant parameter, the threshold should exhibit exponential dependence on applied voltage. Figure~\ref{fig:Ar37_energy_threshold_determination} (right side) demonstrates excellent agreement with the expected behavior through linear fitting in semi-logarithmic coordinates $(V_\mathrm{mesh}, \ln(E_\mathrm{thr}))$.

A caveat in our method is that transfer-region events do not reflect the large Poisson fluctuations that would affect genuine few-electron interactions in the drift region. Consequently, the resolution of a hypothetical 20-eV line would be wider than the narrow peak observed in the transfer region. This would directly affect the shape of the efficiency drop around the threshold. A precise quantification of this effect would require dedicated simulations or a single-electron calibration campaign, but our qualitative conclusion on the achievable threshold remains valid.

%Another potential concern in estimating the threshold could be the presence of backgrounds from single- and few-electron signals, as these sources could impact the smoothness of the transition. We have performed preliminary background runs with the same configuration used in this article, and we observe an increase of single-electron backgrounds at the lowest energies; however, their rate remains low compared with the calibration rate in the runs used in the analysis. We therefore do not consider that these backgrounds affect the threshold estimate presented in this work.

Even considering potential systematic uncertainties from energy scale nonlinearities at ultra-low energies, this analysis demonstrates that TREX-DM is capable of achieving energy thresholds at the level of single-electron ionization energy (26~eV in argon).

\section{Conclusions}
\label{sec:conclusions}

This work demonstrates the successful implementation of a $^{37}$Ar calibration source in the TREX-DM detector, achieving exceptional low-energy threshold performance, crucial for light dark matter searches. The neutron activation of CaO powder at the HiSPANoS facility of CNA in Sevilla provides a practical and effective method for $^{37}$Ar production, yielding sufficient activity for comprehensive detector characterization.

The combined GEM-Micromegas readout system provided preamplification factors of $50-60$, enabling detection of both characteristic $^{37}$Ar emissions. Most notably, the successful observation of the 270~eV L-shell peak represents a significant achievement, as this ultra-low-energy signature is typically challenging to detect in gas-based detectors. The spatial uniformity analysis confirms homogeneous detector response across the active volume, providing a sensitivity map. The systematic threshold analysis reveals remarkable detector sensitivity, with energy thresholds extending down to approximately 20~eV at optimal voltage settings. This performance approaches the fundamental limit imposed by single-electron ionization. Such thresholds have a great impact on the sensitivity projections of TREX-DM to light WIMPs, as discussed in detail in~\cite{GEM_2025}, potentially providing access to unexplored regions of the parameter space.
Future work will focus on using this calibration protocol for precise background characterization studies at low energies. Additionally, efforts are underway to develop more refined calibration techniques to probe this $O(10)$~eV range. A promising approach that has been successfully implemented in the gaseous TPC field~\cite{NEWS-G_2019} is the use of a UV laser to calibrate in the individual electron regime. This would also allow us to accurately study the detection efficiency of our combined GEM-MM detectors to ultra-low-energy events near the energy threshold region.

\acknowledgments

We would like to thank the Servicio General de Apoyo a la Investigación-SAI, Universidad de Zaragoza, for their technical support, and the Micro-Pattern Technologies (MPT) workshop at CERN, where both the Micromegas and the GEM used in this article were manufactured. 
The authors acknowledge support from the European Research Council (ERC) under the European Union’s Horizon 2020 research and innovation programme (ERC-2017-AdG IAXO+, grant agreement No. 788781), from the Agencia Estatal de Investigación (AEI) under the grant agreements PID2021-123879OB-C21 and PID2022-137268NB-C51 funded by MCIN/AEI/10.13039/501100011033/FEDER, as well as funds from “European Union NextGenerationEU/PRTR” (Planes complementarios, Programa de Astrofísica y Física de Altas Energías).

\bibliographystyle{JHEP}
\bibliography{biblio.bib}

\end{document}